\documentclass[ aps, twocolumn, prl,citeautoscript,superscriptaddress,floatfix,notitlepage,groupedaddress,nofootinbib]{revtex4-2}
\usepackage[utf8]{inputenc}
\usepackage{natbib}
\usepackage{graphicx}
\usepackage{amsmath}
\usepackage{amssymb,bm,xspace,soul}
\usepackage{color}
\usepackage[dvipsnames]{xcolor}
\usepackage{anyfontsize} 
\usepackage{units}
\usepackage{times}
\usepackage{setspace}
\usepackage{braket}
\usepackage{empheq}
\usepackage{xcolor}
\usepackage[normalem]{ulem}
\usepackage{slashed}
\usepackage{comment}
\usepackage{adjustbox}
\usepackage{wasysym}
\usepackage[bbgreekl]{mathbbol}
\usepackage{float}
\usepackage[caption = false]{subfig}

\DeclareMathAlphabet{\mathcald}{U}{dutchcal}{m}{n}
\SetMathAlphabet{\mathcald}{bold}{U}{dutchcal}{b}{n}
\DeclareMathAlphabet{\mathalt}{U}{dutchcal}{b}{n}

\DeclareFontFamily{U}{BOONDOX-calo}{\skewchar\font=45 }
\DeclareFontShape{U}{BOONDOX-calo}{m}{n}{
  <-> s*[1.05] BOONDOX-r-calo}{}
\DeclareFontShape{U}{BOONDOX-calo}{b}{n}{
  <-> s*[1.05] BOONDOX-b-calo}{}
\DeclareMathAlphabet{\mathcalb}{U}{BOONDOX-calo}{m}{n}
\SetMathAlphabet{\mathcalb}{bold}{U}{BOONDOX-calo}{b}{n}
\DeclareMathAlphabet{\mathbcalbx}{U}{BOONDOX-calo}{b}{n}

\colorlet{linkColour}{magenta}
\colorlet{citeColour}{OliveGreen}
\colorlet{urlColour}{cyan}

\usepackage[colorlinks=true]{hyperref}
\hypersetup{
    unicode=false,          
    pdftoolbar=true,        
    pdfmenubar=true,        
    pdffitwindow=false,     
    pdfstartview={FitH},    
    pdftitle={Anomalous CFL},    
    pdfauthor={Hart Goldman},     
    pdfsubject={Subject},   
    pdfcreator={Hart Goldman},   
    pdfproducer={}, 
    pdfkeywords={keyword1} {key2} {key3}, 
    pdfnewwindow=true,      
    colorlinks=true,       
    linkcolor=linkColour, 
    citecolor=citeColour,        
    filecolor=linkColour,      
    urlcolor=urlColour           
}

\allowdisplaybreaks

\begin{document}

\title{Zero-field composite Fermi liquid in twisted semiconductor bilayers
}
\author{Hart Goldman$^*$, Aidan P. Reddy$^*$, Nisarga Paul$^*$, and Liang Fu}
\affiliation{Department of Physics, Massachusetts Institute of Technology, Cambridge, MA 02139}
\date{\today}
\def\thefootnote{*}\footnotetext{These authors contributed equally to the development of this work}\def\thefootnote{\arabic{footnote}}

\begin{abstract}
Recent experiments have produced evidence for fractional quantum anomalous Hall (FQAH) states at zero magnetic field in the semiconductor moiré superlattice system $t$MoTe$_2$. Here we argue that a composite fermion description, already a unifying framework for the phenomenology of 2d electron gases at high magnetic fields, provides a similarly powerful perspective in this new context. 
To this end, we present exact diagonalization evidence for composite Fermi liquid states at zero magnetic field in $t$MoTe$_2$ at fillings $n=\frac{1}{2}$ and $n=\frac{3}{4}$. We dub these non-Fermi liquid metals anomalous composite Fermi liquids (ACFLs), and we argue that they play a central organizing role in the FQAH phase diagram.  We proceed to develop a long wavelength theory for this ACFL state 
that 
offers concrete experimental predictions upon doping the composite Fermi sea, including a Jain sequence of FQAH states and a new type of commensurability oscillations originating from the superlattice potential intrinsic to the system. 
\end{abstract}

\maketitle

\newpage 

\textbf{Introduction.} Recently, signatures of fractional quantum anomalous Hall (FQAH) states at zero magnetic field have been observed by optical measurements on twisted bilayer MoTe$_2$ ($t$MoTe$_2$) at fractional fillings of the moir\'e unit cell $n=\frac{2}{3}$ and $\frac{3}{5}$~\cite{Cai2023Signatures}. 
In a separate work, the charge gap of the putative FQAH state at $n=\frac{2}{3}$ was measured~\cite{Zeng2023Integer}.
FQAH states in twisted homobilayers of transition metal dichalcogenides (TMD) were theoretically predicted as a consequence of 
topological moiré bands~\cite{Wu2019Topological}, spontaneous ferromagnetism, and strong correlations~\cite{Devakul2021Magic,Li2021Spontaneous,Crepel2023,Morales2023Pressure}.
These recent observations provide 
new motivation to explore the phenomenology and phase diagram of partially filled Chern bands in $t$MoTe$_2$ and beyond~\cite{Reddy2023, Wang2023Fractional}.

\par Unlike Landau levels, Chern band systems can exhibit competition between incompressible FQAH states ~\cite{Repellin2020Chern, Abouelkomsan2020Particle, Ledwith2020Fractional, Liu2021Gate, Wang2021Exact, Parker2021Field,Neupert2011,Sheng2011,Regnault2011} and more conventional broken symmetry phases enabled by the presence of a periodic lattice structure, such as charge ordered phases~\cite{Kourtis2014,Kourtis2018,Sohal2020} and generalized Wigner crystals~\cite{Regan2020Mott, Xu2020Correlated, Li2021Imaging, Dong22}, or conducting phases~\cite{Barkeshli2012a,Barkeshli2012,Zou2020} 
like Fermi liquids and even superconductors. Exotic quantum critical phases have also been shown to appear in half-filled flat Chern bands~\cite{Song2023}. As a result, the global phase diagram of partially filled Chern bands is potentially much richer than that of Landau levels, calling for systematic study.  

In this work, we focus on the physics of twisted TMD bilayers at even-denominator filling factors, which have not yet received attention. 
We present numerical evidence from continuum model exact diagonalization (ED) calculations of gapless metallic states at filling factors $n=\frac{1}{2}$ and $\frac{3}{4}$.  Remarkably, depending on the twist angle, two types of ferromagnetic metals  with full spin/valley polarization are found. At larger twist angle, the ground state is a Fermi liquid. In contrast, at smaller twist angles where 
strong interaction effect induces odd-denominator FQAH states, we find non-Fermi liquid metals of composite fermions at $n=\frac{1}{2}$ and $\frac{3}{4}$. 
These states share features with the composite Fermi liquid (CFL) at high magnetic fields~\cite{Halperin1993,Halperin2020}, but are ``enriched'' by the underlying moir\'e superlattice. We dub  these zero-field non-Fermi liquid states 
``anomalous composite Fermi liquids'' (ACFLs).

Synonymously with the CFL phases in Landau level systems, we propose the  ACFL as the parent state of the FQAH phase diagram at $B=0$~\cite{Jain1989,Lopez1991,Kivelson1992}. Indeed, based on our ED study, we argue that the prominent FQAH states at $n=\frac{2}{3}$ and $\frac{3}{5}$ in twisted TMD homobilayers are descendants of the ACFL state at $n=\frac{1}{2}$. These states fall along a Jain sequence of FQAH states, which we show emerges by doping the ACFL. 

\begin{figure*}
\centering
\includegraphics[width=\textwidth]{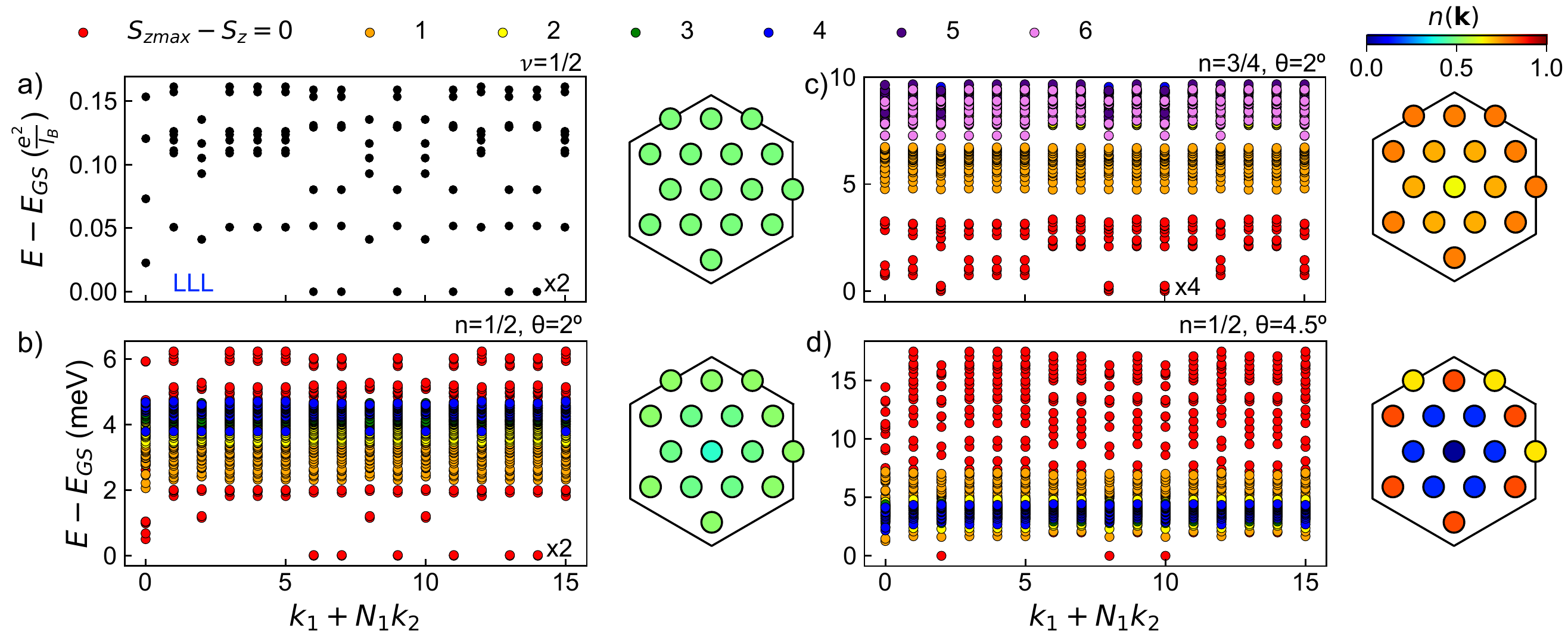}
\caption{\textbf{Zero-field  and ferromagnetic Fermi liquid in semiconductor moiré bands}. Low-lying spectrum as a function of many-body crystal momentum (see the Supplemental Material for elaboration) of the (a) half-filled LLL and (b) lowest moiré band at a twist angle $\theta=2^{\circ}$  with a Coulomb interaction. 
Occupation numbers of the Bloch states averaged over the degenerate ground state manifolds (see main text), $n(\bm{k})\equiv \sum_{i\in GS} \bra{\Psi_i}c^{\dag}_{\bm{k}}c_{\bm{k}}\ket{\Psi_i}/N_{GS}$, are also shown. Full spin polarization is assumed in the LLL whereas all possible $S_z$ sectors are considered in the moiré band. Analogous data at (c) $\theta=2^{\circ}$, $n=\frac{3}{4}$ and (d) $\theta=4.5^{\circ}$, $n=\frac{1}{2}$. In each case the lowest 20 energy levels within each $(\bm{k}, S_z)$ sector are shown. In (b-d) a dielectric constant $\epsilon=10$ is used. 
} \label{fig:mainED}
\end{figure*}

We further reveal the unique phenomenology of the ACFL state itself. Perhaps most strikingly, 
the ACFL resistivity and thermodynamic properties experience \emph{intrinsic commensurability oscillations} as a function of density, $\rho_e$, at $B=0$. This behavior contrasts both with an ordinary Fermi liquid and a CFL in a Landau level~\cite{Smet1998,Smet1999,Willett1999,Kamburov2012,Kamburov2014,Deng2016,Wang2017,Cheung2017,Mitra2019,Shayegan2020,Lu2023}. 
Close to the ACFL state, we find the oscillations are periodic in $1/\delta \rho_e$ and occur at large integers $j$ satisfying
\begin{align}
\frac{1}{\delta \rho_e} \propto \frac{j + \phi}{ k_F\,Q} \,, \label{commensurability}
\end{align}
where $\delta \rho_e\equiv \rho_e-\overline{\rho}$ is the doping density from half filling, $Q$ is the moir\'{e} superlattice wave vector,  $k_F=\sqrt{4\pi \overline{\rho}}$ is the composite Fermi wave vector, and $\phi$ is a phase shift.   
For the ACFL state at a particular even-denominator filling fraction such as $\overline{n}=\frac{1}{2}$, $k_F$ is proportional to $Q$, meaning that the doping density, $\delta \rho_e$, associated with the commensurability oscillation is inversely proportional to the moir\'e unit cell area. As a result, the corresponding filling fraction, $n=\overline{n}+\delta n$, is universal.  

Eq.~(\ref{commensurability}) is both a consequence of the attachment of flux to charge in the ACFL -- which causes the composite fermions to feel an effective magnetic field upon doping 
-- and the system's intrinstic moiré potential. 
In contrast, commensurability oscillations in CFLs in Landau level systems require an \emph{externally supplied} periodic potential. In addition to commensurability oscillations, a  distinguishing feature of a CFL (at zero or finite field) is a large DC Hall angle, ${\theta_H=\arctan(\sigma_{xy}/\sigma_{xx})}$, which approaches $\pi/2$ in the clean limit~\cite{Halperin1993}. 

CFL phases are expected to exhibit non-Fermi liquid observable features, some of which may be accessible as new platforms are developed realizing the ACFL. For example, the 
thermodynamic entropy of the ACFL state can be measured from the change in chemical potential with temperature through a Maxwell relation ~\cite{Saito2020Isospin, Li2021Continuous}. In the clean limit, because gauge fluctuations lead to a logarithmic mass enhancement of composite fermions, the entropy of ACFL state should also be enhanced~\cite{Sheng2020Thermoelectric}, ${s(T)/T\sim m_*(T)\sim\log T}$~\cite{Halperin1993}, compared to  the linear temperature dependence of an ordinary Fermi liquid, 
${s(T)/T\sim\,}$constant. In systems where the electronic Coulomb interaction is screened by a nearby metallic gate, this enhancement becomes a power law, ${s(T)/T\sim T^{-1/3}}$. 




\textbf{Motivation.} In ordinary Landau level quantum Hall systems, the existence of a CFL phase can be understood through flux attachment. At even-denominator filling ${\nu=2\pi \rho_e/B=\frac{1}{2q}}$, where $q$ is an integer, $\rho_e$ is the electron density, and $B$ is the external magnetic field, attaching $2q$ flux quanta to each electron completely screens the magnetic field, leading to an effective system of composite fermions in effective magnetic field $b_*=B-2q(2\pi\rho_e)=0$. As a result, the composite fermions form a Fermi surface strongly coupled to an emergent gauge field~\cite{Halperin1993}. Upon 
doping away from $\nu=\frac{1}{2q}$, the composite fermions feel a nonvanishing magnetic field and fill Landau levels, leading to the Jain sequence of observed fractional quantum Hall phases, 
\begin{align} 
\label{eq: nuJain}
\nu_{\mathrm{Jain}}&=\frac{p}{2q\,p-1}\,,
\end{align}
where $p$ is the number of filled composite fermion Landau levels~\cite{Jain1989,Lopez1991}. 

A similar picture 
should be applicable to twisted TMD bilayers in the absence of a physical magnetic field, when (1) interactions spontaneously drive all carriers into the Chern band of one valley; (2) the Coulomb Hamiltonian projected to the Chern band sufficiently resembles that of the lowest Landau level (LLL); and (3) the band dispersion is small 
relative to the system's characteristic interaction energy scale $\sim e^2/(\epsilon a_M)$.   
When these conditions are satisfied, the problem can be approximately mapped to that of a partially filled Landau level. 
One might therefore expect that any of the quantum Hall phases in a Landau level at filling $\nu$ should be possible in a flat $C=1$ band at the same filling. 
The challenge is to find situations in which such physics succeeds over other phases that are not possible in Landau levels. Hence, once a material is known to exhibit the FQAH effect, it is natural to anticipate that other essential features of the fractional quantum Hall phase diagram also occur in the same material, such as the convergence of Jain sequence FQAH phases into a metallic CFL at half-filling. 


\textbf{Numerical evidence for ACFL.} 
We now provide numerical evidence for 
the ACFL in $t$MoTe$_2$. The 
nontrivial layer pseudospin structure of this system's Bloch wavefunctions endows its moiré bands with topological character~\cite{Wu2019Topological}. In particular, the first moiré valence band in each valley has $|C|=1$, with opposite signs in opposite valleys due to time-reversal symmetry. 
We study the continuum model of $t$MoTe$_2$ 
with Coulomb interaction $U(r)=\frac{e^2}{\epsilon r}$ projected to the lowest moiré band, using finite size ED with torus geometry \cite{Reddy2023}. 
Further details of the model and methodology are provided in the Supplemental Material.

To establish a benchmark 
for 
a CFL on a finite-size torus, we show the low-lying many-body spectrum of the half-filled LLL on a torus
with 16 flux quanta 
in Fig.~\ref{fig:mainED}(a). 
Given our system geometry, the spectrum features 12 exactly degenerate ground states with 2 in each of 6 momentum sectors. The momentum quantum numbers of the degenerate ground states reflect 
the most compact possible composite Fermi sea configurations~\cite{Wang2021Exact}. There are 6 such configurations -- one composite fermion is accounted for by occupying the state at the center of the Brillouin zone, 6 more by occupying the set of closest points
, and the final one by occupying any one of the 6 next closest points. The additional factor of 2 in the overall ground state degeneracy is enforced by the non-commuting center-of-mass magnetic translations 
~\cite{haldane1985many}. We also show the momentum space occupation numbers of electron Bloch states averaged over the ground state manifold. The Bloch state occupation is uniform despite the presence of a composite Fermi sea. 

Next, in Fig.~\ref{fig:mainED}(b), we show the many-body spectrum of the $\theta=2^{\circ}$ lowest $t$MoTe$_2$ band at filling $n=\frac{1}{2}$ on the corresponding 16-unit-cell torus across all possible $S_z \geq 0$ sectors. First, we observe that the lowest-lying states have $S_z=S_{z,\mathrm{max}}=4$, indicating spontaneous, full spin/valley polarization. Moreover, these states have the same momentum quantum numbers as their partners in the LLL, providing evidence for a composite Fermi sea. 
The momentum space occupation numbers are nearly uniform as in the LLL, demonstrating that the system is \emph{not} a ferromagnetic Fermi liquid. In Fig.~\ref{fig:mainED}(c), we show an ED spectrum at $n=\frac{3}{4}$ that exhibits full valley polarization and similarly resembles 
its partner in the LLL, which is shown in the Supplemental Material.

In Fig.~\ref{fig:mainED}(d), we contrast these findings with $n=\frac{1}{2}$ at a larger twist angle $\theta=4.5^{\circ}$. Here, the lowest-energy states are still fully spin/valley polarized, but their many-body momenta are those expected from simply occupying the moiré band Bloch states with lowest energy, indicating a Fermi liquid phase. Moreover, the 
Bloch state occupation numbers in Fig.~\ref{fig:mainED}(c) exhibit a sharp drop across the Fermi surface expected for non-interacting, spin-polarized holes. 


\textbf{Effective theory of the ACFL.}  With this motivation, we propose a long wavelength effective theory of the ACFL in a Chern band at half-filling, $n=\frac{1}{2}$,
\begin{align}
\label{eq: ACFL}
\mathcal{L}_{\mathrm{ACFL}}&=\psi^\dagger\Big[i\partial_t+a_t+A_t+\mathcal{V}(\boldsymbol{x})\Big]\,\psi\nonumber\\
&\qquad-\frac{1}{2m_*}\left| (i\partial_i+a_i+A_i)\psi\right|^2-V(\rho_e)\nonumber\\
&\qquad-\frac{1}{2}\frac{1}{4\pi}\,\varepsilon_{\mu\nu\lambda}\,a_\mu\partial_\nu a_\lambda-a_t\,\overline\rho\,.
\end{align}
Here $\psi$ is the composite fermion field, $m_*$ is an effective mass; $V(\rho_e)$ is the density-density interaction potential; 
${a_\mu=(a_t,a_x,a_y)}$ is a fluctuating Chern-Simons statistical gauge field; and ${A_\mu=(A_t,A_x,A_y)}$ is the background electromagnetic gauge field. We denote the value of the charge density at half filling by $\overline\rho$, such that the charge per unit cell is ${\overline{n}\equiv\overline\rho\times(\textrm{unit cell area})=\frac{1}{2}}$. Importantly, although we focus on $n=\frac{1}{2}$ here, the theory for the ACFL at $n=\frac{3}{4}$ is easily obtained by attaching $4$ flux quanta and acting with a particle-hole transformation (subtracting a filled $C=1$ band). We expect its universal properties to be essentially the same as the ACFL at $n=\frac{1}{2}$. In the Supplemental Material, we discuss how Eq.~\eqref{eq: ACFL} can be constructed from a parton mean field construction along similar lines to Ref.~\cite{Barkeshli2012}, which also considered ACFL type phases in flat Chern bands (we note here that the Supplemental Material also includes the additional Refs.~\cite{Wu2013Bloch,Bernevig2012,Jain1989a,Wen1992,Vaezi2011,Lu2012,Sohal2018,MacDonald1983,Lu2020,Manjunath2021,Pippard1962,Pippard1964,Girvin1984,Son2015,Goldman2018,Kumar2018,Kumar2019a}).

The theory in Eq.~\eqref{eq: ACFL} closely resembles the Halperin-Lee-Read (HLR) theory of half-filled Landau levels~\cite{Halperin1993}. This is not surprising: the $|C|=1$ Chern band in twisted TMD bilayers  can be thought of as carrying an emergent magnetic flux of one flux quantum per unit cell, which arises from the skrymion lattice configuration of ``Zeeman'' field acting on the layer pseudospin ~\cite{Wu2019Topological}. Nevertheless, there are two major differences with the standard HLR theory for a half-filled Landau level. The first is the final term, which alters the usual flux attachment constraint. Using the equation of motion for $a_t$,
\begin{align}
\label{eq: modified flux attachment}
\rho_e=\overline\rho+\frac{1}{2}\frac{1}{2\pi}(\boldsymbol{\nabla}\times\boldsymbol{a})\,,
\end{align}
where we have used the fact that the physical electron density consides with that of the composite fermions, ${\rho_e=\delta\mathcal{L}_{\mathrm{ACFL}}/\delta A_t=\psi^\dagger\psi}$, and boldface denotes spatial vectors. At half-filling of the Chern band, $n=\overline{n}=\frac{1}{2}$, meaning that by Eq.~\eqref{eq: modified flux attachment} the gauge flux per unit cell must vanish, and the composite fermions form a Fermi surface (as above, we use $n$ to denote charge per unit cell). 

The second difference is that Eq.~\eqref{eq: ACFL} \emph{includes the effect of the moir\'{e} superlattice}, in the form of a periodic scalar potential, 
\begin{align}
\mathcal{V}(\boldsymbol{x})=\mathcal{V}_0\sum_{n=1}^3\cos\left(\boldsymbol{Q}_n\cdot\boldsymbol{x}\right)\,,
\end{align}
where $\boldsymbol{Q}_n$ are the moir\'{e} superlattice wave vectors (see Supplemental Material). The full scalar potential felt by the composite fermions is therefore $\mathcal{V}(\boldsymbol{x})+A_t$, where $A_t$ includes any additional probe fields. We will see that the presence of this term leads to commensurability oscillations which are unique to the ACFL. 

\begin{figure}
\centering
\includegraphics[width=\columnwidth]{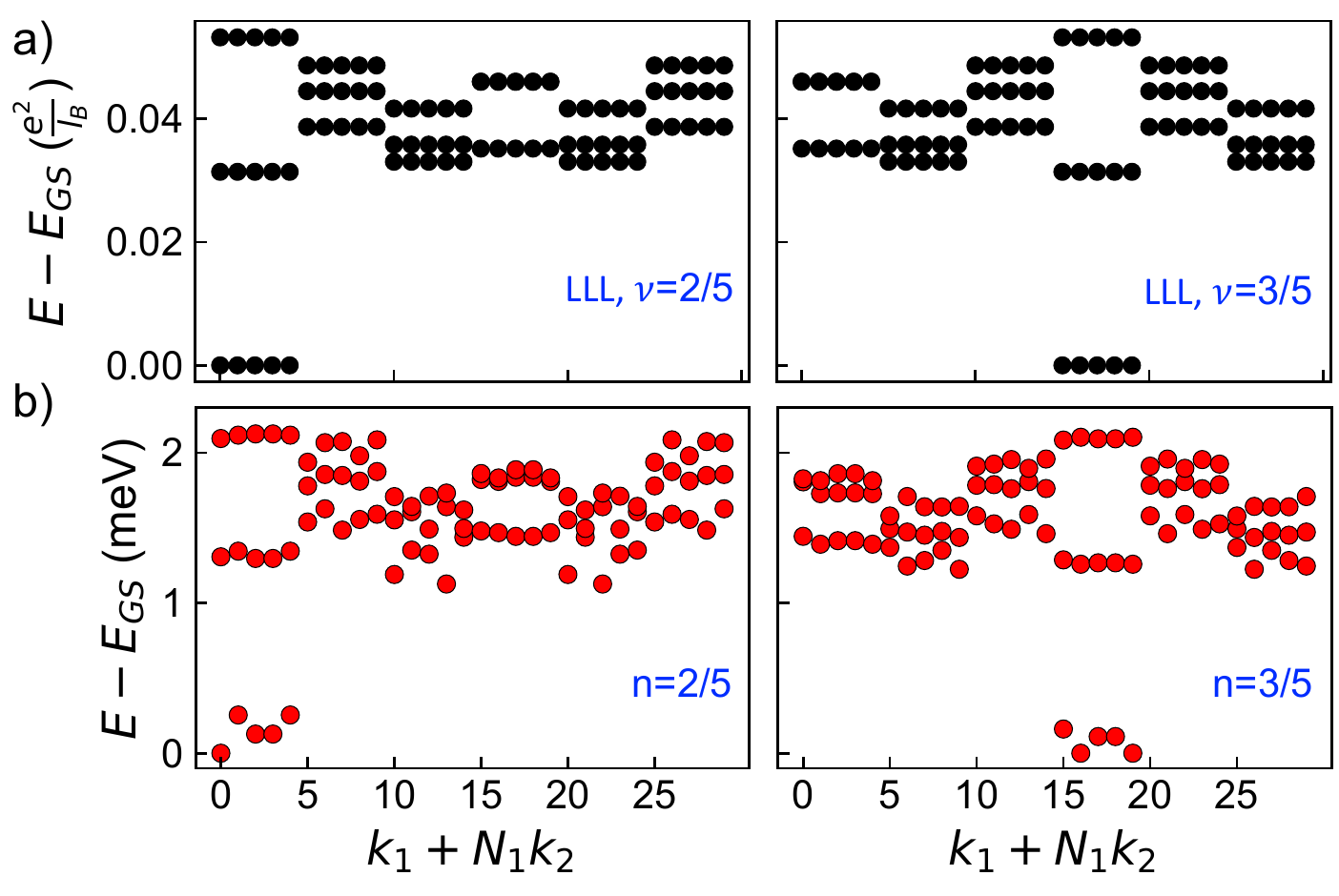}
\caption{\textbf{Jain sequence states at $n=\frac{2}{5}$, $\frac{3}{5}$}. (a) Low-lying spectrum of the LLL with Coulomb interaction on a torus with 30 flux quanta at $\nu=\frac{2}{5}$ and $\frac{3}{5}$. (b) Corresponding data for the lowest moiré band at $\theta=2^{\circ}$, $\epsilon=10$. The moiré ground state manifold has the same momentum quantum numbers and approximate fivefold topological ground state degeneracy as the LLL. In each case 
the lowest 3 fully spin polarized states in each momentum sector are shown.}\label{fig:jainsequence}
\end{figure}

Although the theory in Eq.~\eqref{eq: ACFL} should correctly reproduce long wavelength, universal observable properties, we emphasize that this theory is not meant to completely incorporate microscopic details. For example, it does not give the correct algebra of density operators, nor does it incorporate the composite fermion dipole moment. Rather, we expect that should a complete, band-projected theory be constructed, then Eq.~\eqref{eq: ACFL} could be understood as its long wavelength limit. For recent efforts to develop band-projected composite fermion theories in the context of the LLL, see Refs.~\cite{Zhihuan2020,Zhihuan2022,Goldman2022,Milica2023}.     

\begin{figure}
    \centering
    \includegraphics[width=0.5\linewidth]{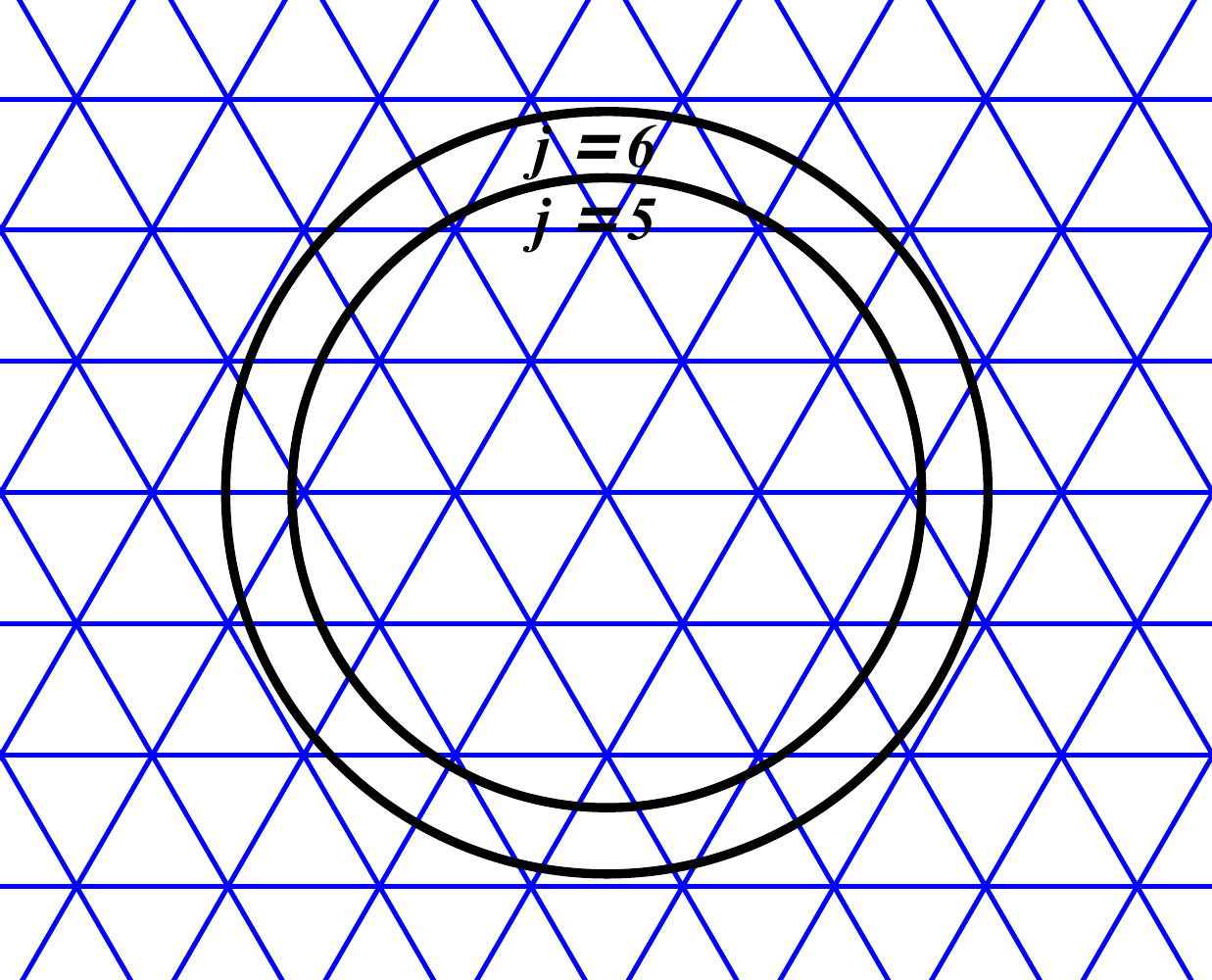}
\caption{\textbf{Commensurability oscillations. } Schematic of cyclotron orbits at special commensurate values. For $j\geq5$, $\frac12 < \rho_e<\frac35$.}
\label{fig:magnetic}
\end{figure}

\textbf{FQAH sequence.} Doping away from half-filling by tuning charge density or applied magnetic field causes the composite fermions to feel a net magnetic field and fill Landau levels. As a result, we can immediately predict a Jain sequence of FQAH states in $t$MoTe$_2$ 
corresponding to integer quantum Hall states of composite fermions. Like fractional Chern insulator states developed earlier~\cite{Moller2009,Neupert2011,Sheng2011,Regnault2011,Tang2011}, these FQAH phases are topological orders enriched with (super)lattice symmetry. Say that the composite fermions fill $p$ Landau levels,
\begin{align}
\label{eq: CF filling}
\nu_\psi=2\pi\frac{\langle\psi^\dagger\psi\rangle}{b_*}=p&,\qquad b_*=\boldsymbol{\nabla}\times(\boldsymbol{a}+\boldsymbol{A})\,,
\end{align}
where $b_*$ is the total magnetic field felt by the composite fermions. Combining the flux attachment constraint, Eq.~\eqref{eq: modified flux attachment}, with Eq.~\eqref{eq: CF filling}, we can relate the electron density to the applied magnetic field, ${B=\boldsymbol{\nabla}\times\boldsymbol{A}}$,
\begin{align}
\label{Eq: Jain density}
\rho_e(B)&=\frac{p}{2p-1}\left(2\overline{\rho}-\frac{B}{2\pi}\right)\,.
\end{align}
The Streda formula then implies that one will measure Landau fans extending to $B=0$, with slopes that fall on the Jain sequence (see Fig.~\ref{fig:magnetic}),
\begin{align} 
\frac{d\rho_e}{dB}=\sigma_{xy}=-\frac{p}{2p-1}\,.
\end{align}
For the FQAH sequence proximate to $n=\frac{3}{4}$, one obtains FQAH states on the sequence, $n=1-\frac{p}{4p-1}$. 

It is instructive to multiply Eq.~\eqref{Eq: Jain density} on both sides by the superlattice unit cell area to obtain the simple expression,
\begin{align}
n=\frac{p}{2p-1}\left(1-n_\Phi\right)\,,
\end{align}
where $n_\Phi$ is the flux per unit cell. The FQAH Jain sequence includes 
the observed state at filling $\frac{2}{3}$ in $t$MoTe$_2$, which has also been studied numerically~\cite{Reddy2023, Wang2023Fractional}. In Fig.~\ref{fig:jainsequence} we present ED evidence 
for additional Jain FQAH states in $t$MoTe$_2$ at $n=\frac{2}{5}$ and its particle-hole conjugate at $n=\frac{3}{5}$. 

\textbf{Intrinsic commensurability oscillations at zero field. } We now explore the host of phenomena arising from an interplay of flux attachment with the presence of the moir\'{e} superlattice potential. Indeed, due to the periodic modulation intrinsic to moir\'e materials, we find that both commensurability oscillations~\cite{Weiss1989,Gerhardts1989,Winkler1989,Weiss1990,Gerhardts1996} and Hofstadter subgaps can be accessed by tuning density \textit{alone}. \par 
Commensurability oscillations occur when the cyclotron radius and the modulation period are commensurate. More precisely, magnetoresistance minima and compressibility maxima are expected when a system in a spatially modulated potential with wave vector $Q$ satisfies the electronic flat band conditions, 
\begin{equation}
    2k_F Q\,\ell^2 = 2\pi(j + \phi)\,,
\end{equation}
where $j$ is a positive integer, $\ell$ is the effective magnetic length felt by the electric charges, and $\phi$ is a phase shift. This condition is derived in the Supplemental Material 
from both perturbative and semiclassical approaches.

Again focusing on the $n=\frac{1}{2}$ ACFL, if the system is doped from half-filling by density, $\delta \rho_e=\rho_e-\overline{\rho}$, the composite fermions feel a magnetic field $b_* = 4\pi\, \delta \rho_e$ by the flux attachment constraint in Eq.~\eqref{eq: modified flux attachment}. As in the HLR approach, the composite Fermi wave vector of the ACFL described by Eq.~\eqref{eq: ACFL} is set by the electric charge density, ${k_F = \sqrt{4\pi \overline{\rho}} }$. We then find that commensurability oscillations occur at densities
\begin{equation}\label{eq:weiss}
    \delta \rho_e = \frac{k_F\,Q}{(2\pi)^2} \frac{1}{j + \phi}
\end{equation}
 at large $j$ where $\delta \rho_e$ is small compared to $\overline{\rho}$. 
For the $n=1/2$ ACFL, we have $k_F Q = (4\pi^{3/2}/\sqrt[4]{3})/A$ ($A$ is the moir\'e unit cell area) so that the oscillations correspond to filling fractions, ${n = \frac12 +\delta n}$, where 
 \begin{equation}
\delta n
 = \frac{1}{\sqrt[4]{3}\sqrt{\pi}}\frac{1}{j+\phi}\approx \frac{0.43}{j+\phi}\,.    
 \end{equation}
Close to $n=1/2$, the oscillations therefore have period $\Delta(1/\delta n)\approx 2.3$. 
 
 We emphasize that in the ACFL these commensurability oscillations occur \emph{in the absence of any external magnetic field.} 
 They coexist with SdH oscillations coming from filling integer Landau levels of composite fermions, which realize Jain states in a clean system. 
Additionally, tuning density and external field together allows full access of the magnetic spectrum of composite fermions near  half-filling of the Chern band. 


\textbf{Discussion.}
Starting from a band-projected continuum model for $t$MoTe$_2$, we have presented exact diagonalization evidence for compressible, non-Fermi liquid states at zero magnetic field, which we dub the anomalous composite Fermi liquid. Much as in conventional fractional quantum Hall systems, we argue that the ACFL picture offers a powerful organizing perspective for understanding fractional quantum anomalous Hall states. Indeed, all of the states for which there currently exists theoretical or experimental evidence fall on the celebrated Jain sequence. 
We furthermore developed an effective theory capturing the universal properties of the ACFL offering concrete, observable signatures, including commensurability oscillations and Jain sequence FQAH states themselves.

Interestingly, in the recent experiment on $t$MoTe$_2$~\cite{Cai2023Signatures}, 
the coercive field is found to be enhanced at $n=\frac{3}{4}$, in addition to $n=\frac{2}{3}$. Unlike the $n=\frac{2}{3}$ state, which is incompressible, the $n=\frac{3}{4}$ state appears to be compressible. Our theory offers a potential explanation of the observed $n=\frac{3}{4}$ state as the ACFL. 


Finally, the emergence of an ACFL state suggests the possibility of new quantum phase transitions not possible at finite field. For example, while our analysis here focuses on zero displacement field, it should be possible to induce a phase transition between ACFL and ferromagnetic Fermi liquid phases by tuning displacement field. Continuous transitions between CFLs and Fermi liquids have been proposed in the past~\cite{Barkeshli2012,Zou2020}. We leave study of such phase 
transitions to future work.


\emph{Note added.} 
Recently, an independent work by Dong \emph{et al.}~\cite{Dong2023} appeared, which has some overlapping conclusions with the present study. 

\textbf{Acknowledgements.} We are grateful to Xiaodong Xu and Ady Stern for interesting discussions. HG also thanks Jennifer Cano, Mike Mulligan, Sri Raghu, T. Senthil, Raman Sohal, Alex Thomson, and Cenke Xu for conversations on related topics. This work was supported by the Air Force Office of Scientific Research (AFOSR) under award FA9550-22-1-0432 and the David and Lucile Packard Foundation. 
HG is supported by the Gordon and Betty Moore Foundation
EPiQS Initiative through Grant No. GBMF8684 at the Massachusetts Institute of Technology.  The authors acknowledge the MIT SuperCloud and Lincoln Laboratory Supercomputing Center for providing HPC resources that have contributed to the research results reported within this paper.


%

\end{document}


\onecolumngrid

\title{Supplemental material: \\ Zero-field composite Fermi liquid in twisted semiconductor bilayers }
\author{Hart Goldman}
\author{Aidan P. Reddy}
\author{Nisarga Paul}
\author{Liang Fu}
\affiliation{Department of Physics, Massachusetts Institute of Technology, Cambridge, MA 02139}
\maketitle

\section{Exact diagonalization methodology and extended data}

Our exact diagonalization calculation starts with the single-particle continuum model for AA-stacked, K-valley TMD moiré homobilayers:
\begin{eqnarray}\label{eq:H0}
H_0 = \sum_{\sigma=\uparrow, \downarrow} \int d {\bm r} \; \psi_{\sigma}^\dagger(\bm{r}) \mathcal{H}_{\sigma} \psi_{\sigma}(\bm{r}).
\end{eqnarray}
Here $\mathcal{H}_{\sigma}$ is a $2\times2$ matrix in layer pseudospin,
\begin{align}\label{eq:CMHam}
  \mathcal{H}_{\uparrow} = \begin{pmatrix}
  \frac{\hbar^2(- i \nabla - \kappa_+)^2}{2m} + V_1(\bm{r}) & t(\bm{r}) \\
  t^{\dag}(\bm{r}) & \frac{\hbar^2( - i\nabla - \kappa_-)^2}{2m} + V_2(\bm{r})
  \end{pmatrix}
\end{align}
where
\begin{align}
V_l(\bm{r})=-2V\sum_{i=1,3,5}\cos(\bm{g}_i\cdot\bm{r}+(-1)^l\phi),
\end{align}
\begin{align}
  t(\bm{r}) = w(1+e^{i\bm{g}_2\cdot\bm{r}}+e^{i\bm{g}_3\cdot{\bm{r}}}),
\end{align}
and $\bm{g}_i=\frac{4\pi}{\sqrt{3}a_M}(\cos\frac{\pi(i-1)}{3},\sin\frac{\pi(i-1)}{3})$, $\kappa_- = \frac{\bm g_1 +\bm g_6}{3}$, $\kappa_+ = \frac{\bm g_1 +\bm g_2}{3}$. $\mathcal{H}_{\downarrow}$ is fixed by $\mathcal{H}_{\uparrow}$ and the time-reversal symmetry condition $\mathcal{T}H_0\mathcal{T}^{-1}=H_0$. Due to large spin-orbit splitting at the monolayer valence band maxima, the spin and valley are locked at low energies into a single ``spin" quantum number indexed by $\sigma$. This model has been derived and its rich single-particle properties discussed elsewhere~\cite{Wu2019Topological, Devakul2021Magic, Reddy2023}. We work under a particle-hole transformation $c_{\bm{k}\uparrow} \rightarrow c^{\dag}_{\bm{k}\uparrow}$ so that the spectrum is bounded from below.

We study the model's many-body physics by adding a Coulomb interaction $V(r) = \frac{e^2}{\epsilon r}$ and projecting to the lowest moiré band, yielding the projected Hamiltonian
\begin{align}\label{eq:projham}
\begin{split}
    \tilde{H} &= \sum_{\bm{k},\sigma}\varepsilon_{\bm{k}\sigma}c^{\dag}_{\bm{k}\sigma}c_{\bm{k}\sigma} + \frac{1}{2}\sum_{\bm{k}'\bm{p}'\bm{k}\bm{p},\sigma\sigma'}V_{\bm{k}'\bm{p}'\bm{k}\bm{p};\sigma\sigma'}c^{\dag}_{\bm{k}'\sigma}c^{\dag}_{\bm{p}'\sigma'}c_{\bm{p}\sigma'}c_{\bm{k}\sigma}
  \end{split}
\end{align}
where $c^{\dag}_{\bm{k}\sigma}$ creates a hole with moiré crystal momentum $\bm{k}$ spin/valley quantum number $\sigma$ with band energy $\varepsilon_{\bm{k}\sigma}$. $V_{\bm{k}'\bm{p}'\bm{k}\bm{p};\sigma\sigma'} \equiv \bra{\bm{k}'\sigma;\bm{p}'\sigma'}\hat{V}\ket{\bm{k} \sigma;\bm{p} \sigma'}$ are the Bloch state interaction matrix elements. This method can be viewed as a variational approximation of the ground and low-lying excited states with the Fock space of the lowest moiré band as the variational Hilbert subspace. This variational approximation is designed to be exact in the absence of interactions and will be generically inexact in their presence.


The torus geometries used in the exact diagonalization calculations shown in this work are defined by the boundary condition wavevectors $\bm{L}_1,\bm{L}_2$. The 16-unit-cell cluster is defined by $\bm{L}^{(16)}_i=4\bm{a}_i$ and the $30$-unit-cell cluster by $\bm{L}^{(30)}_1=5\bm{a}_1$, $\bm{L}^{(30)}_2=6\bm{a}_2$ where $\bm{a}_1=\lambda(\frac{1}{2},\frac{\sqrt{3}}{2})$, $\bm{a}_2=\lambda(0,1)$. $\lambda=a_M$ in the case of the moiré system and $\lambda=\frac{\sqrt{2\pi\ell^2}}{\sqrt{\sqrt{3}/2}}$ in the case of the LLL where $\ell$ is the magnetic length such that the moiré and magnetic unit cells correspond.

In the moiré case, the many-body crystal momentum quantum number $\bm{k}$ of a given many-body state $\ket{\Psi}$ is defined as $\bm{T}^{CM}_{\bm{a}_i}\ket{\Psi}=e^{i\bm{k}\cdot\bm{a}_i}\ket{\Psi}$ where $T^{CM}_{\bm{d}}=\prod_i e^{i\bm{p}_i\cdot\bm{d}}$ is an ordinary center-of-mass translation operator where the index $i$ labels particles. $\bm{k}=k_1\bm{T}_1+k_2\bm{T}_2$ where $\bm{L}_i=N_i\bm{a}_i$, and $\bm{T}_1=\frac{2\pi\epsilon_{ij}\bm{L}_j\times\hat{z}}{|\bm{L}_1\times\bm{L}_2|}$. In the case of the LLL, we define the many-body crystal momentum quantum number as the eigenvalue under a center-of-mass \emph{magnetic} translation by a primitive lattice vector
${t^{CM}_{\bm{a}_i}\ket{\Psi} = e^{i\bm{k}\cdot\bm{a}_i} \ket{\Psi}}$ where $t^{CM}_{\bm{d}}=\prod_i t^i_{\bm{d}}$ is the center-of-mass magnetic translation operator. Ref.~\cite{Wu2013Bloch} explains how to construct set of the LLL single-particle states on a torus that transform like Bloch states under \emph{magnetic} translations, $t_{\bm{a}_i}\ket{\psi_{\bm{k}}}=e^{i\bm{k}\cdot \bm{a}_i}\ket{\psi_{\bm{k}}}$. A Fock state constructed from a set of these single-particle states has center-of-mass magnetic crystal momentum equal to the sum of the magnetic crystal momenta of those single-particle states, analogously to the zero-$B$ moiré case. We note that this approach does not exploit the full many-body translation symmetries present in a Landau level (originally presented in Ref.~\cite{haldane1985many} and explained more pedagogically in Ref.~\cite{Bernevig2012}), but is convenient in this case because it allows for direct comparison with the moiré system.

We use the continuum model parameters reported in Ref.~\cite{Reddy2023} for $t$MoTe$_2$. A more detailed account of our continuum model exact diagonalization methods is presented in the Supplemental Material of Ref. \cite{Reddy2023}.

\begin{figure}[H]
\centering
\includegraphics[width=0.5\columnwidth]{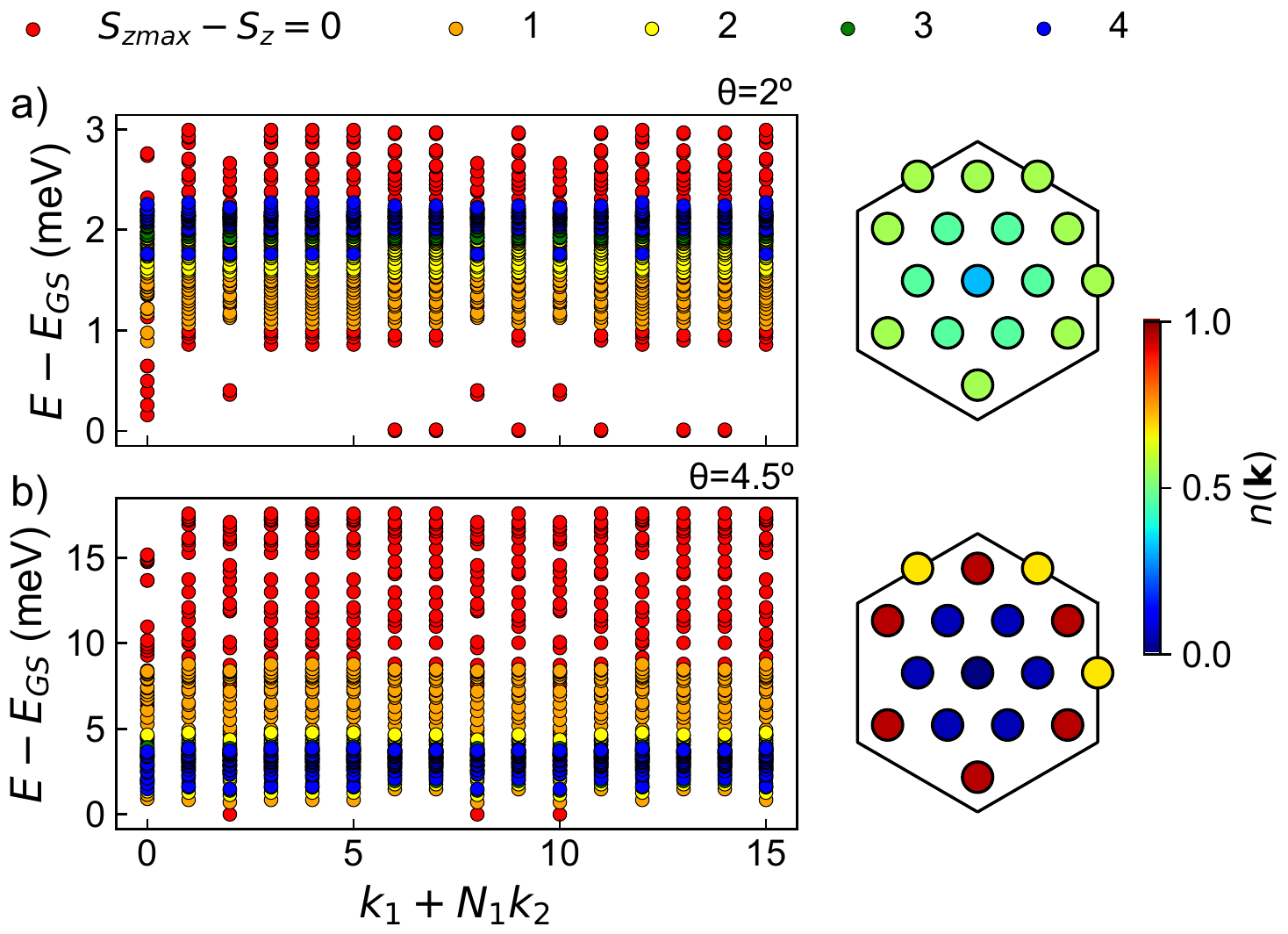}
\caption{\textbf{Robustness of AFCL and ferromagnetic metal against weaker interaction}. Analogous data to that shown in Fig.~1 of the main text at $n=\frac{1}{2}$ but in the presence of a weaker Coulomb interaction $\epsilon=20$.  }\label{fig:halffilledweakint}
\end{figure}

\begin{figure}[H]
\centering
\includegraphics[width=0.51\columnwidth]{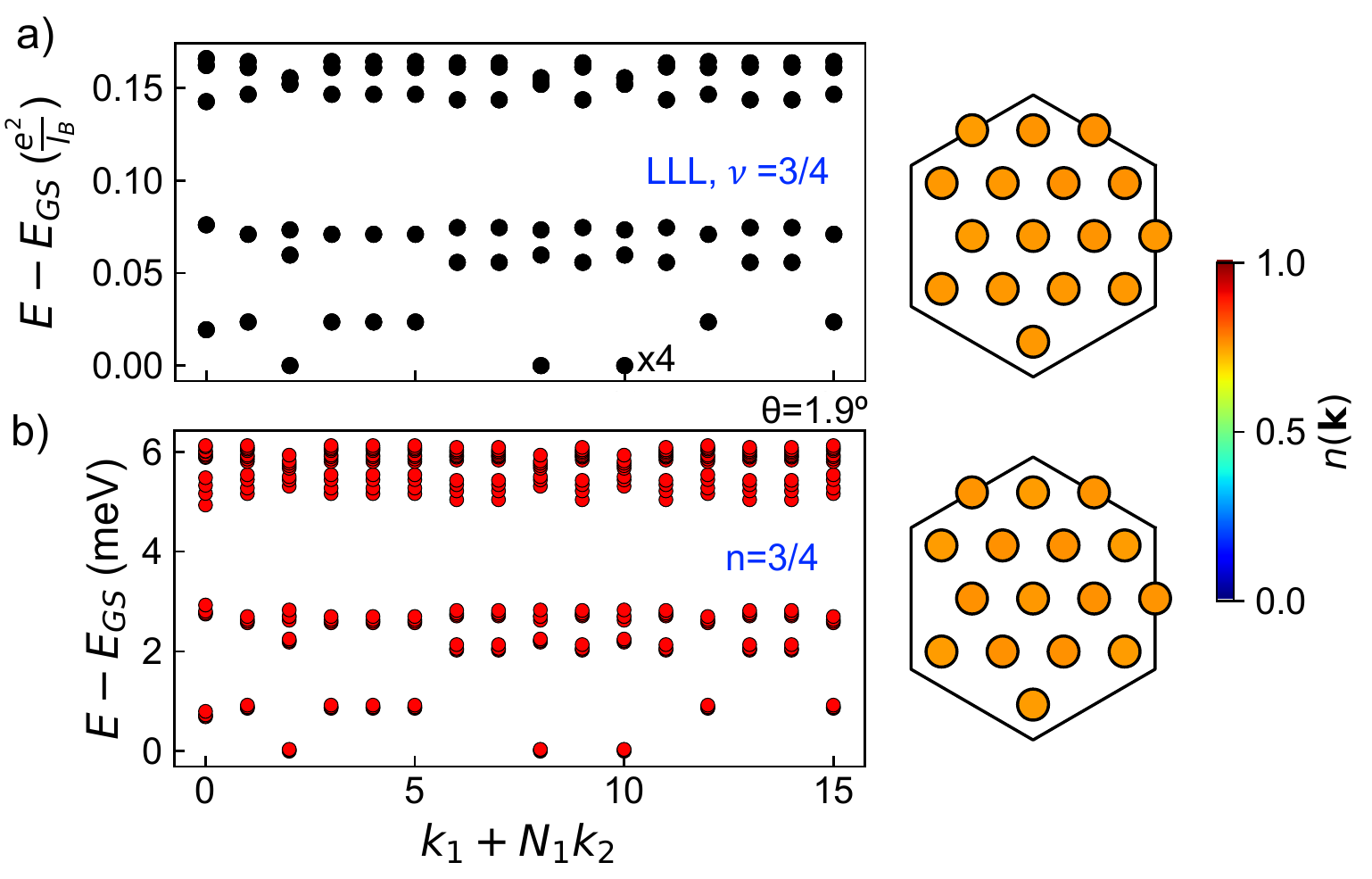}
\caption{\textbf{ACFL at $n=\frac{3}{4}$} (a) Spectrum of fully spin-polarized electrons with a Coulomb interaction at $\nu=\frac{3}{4}$ in the LLL. Its low-lying states have the same many-body momenta and degeneracy as in the lowest moiré band shown in Fig.~1(c) of the main text. (b) Corresponding spectrum in the lowest moiré band at $\theta=1.9^{\circ}$, $\epsilon=10$, resembling the LLL even more closely than $\theta=2^{\circ}$ shown in Fig.~1(c) of the main text. On the right hand side are the occupation numbers of the single-particle crystal momentum states averaged over the 12-fold nearly degenerate ground state manifold, demonstrating the absence of an electronic Fermi surface. The lowest 20 energy levels in each momentum sector are shown.}\label{fig:threequarters}
\end{figure}

In Fig. \ref{fig:halffilledweakint}, we provide evidence that the same qualitative results for the composite Fermi liquid  and ferromagnetic metal at $n=\frac{1}{2}$ occur in the presence of weaker Coulomb interaction $\epsilon=20$. In Fig \ref{fig:threequarters}(a), we show the ED spectrum of the LLL at $\nu=\frac{3}{4}$ with which Fig.~1(c) can be compared. Finally, in Fig. \ref{fig:threequarters}(b), we show that the spectrum of the moiré AFCL at $n=\frac{3}{4}$, $\theta=1.9^{\circ}$ resembles its LLL cousin even more closely than at $\theta=2^{\circ}$ shown in Fig~1(c) of the main text. 

\section{Parton construction of the ACFL effective theory}

Despite the absence of external magnetic flux in the Chern band context, much of the physics of flux attachment persists if we adopt a perspective where the physical electrons, denoted $c$, are fractionalized into emergent parton degrees of freedom, 
\begin{align}
\label{eq: partons}
c_x=\phi_x\, \psi_x\,.
\end{align}
Here $\phi$ is taken to be a bosonic ``holon'' that carries the physical electric charge, while $\psi$ is a neutral fermionic ``spinon.'' Using this parton description allows for a simple extension of the intuition of flux attachment to the Chern band setting~\cite{Jain1989a,Wen1992,Vaezi2011,Lu2012,McGreevy2012,Barkeshli2012a,Barkeshli2012}. However, we note recent efforts to construct lattice flux attachment procedures for constructing FQAH and fractional Chern insulator phases~\cite{Sohal2018,Sohal2020}.

The above parton decomposition has a local gauge redundancy, under which $\psi_x\rightarrow e^{i\theta_x}\psi_x$ and $\phi_x\rightarrow e^{-i\theta_x}\phi_x$, meaning that both must also couple to an emergent U$(1)$ gauge field which we denote $a_\mu=(a_t,a_x,a_y)$. In summary, $\phi$ couples to the combination $A_\mu-a_\mu$, where $A_\mu=(A_t,A_x,A_y)$ is the physical (background) electromagnetic gauge field, while $\psi$ couples to $a_\mu$ alone. We will assume that the system is in a regime where gauge fluctuations do not confine the partons.

Previously, Refs.~\cite{Barkeshli2012,Zou2020} applied the parton decomposition in Eq.~\ref{eq: partons} to study ACFL phases starting from lattice tight binding models. Here, with an eye toward moir\'{e} systems, we approach the emergence of the ACFL from a more long wavelength perspective. We treat the long-wavelength moir\'{e} superlattice as a periodic electric scalar potential, $A_t=\mathcal{V}(x)$. Then, neglecting gauge fluctuations, $\langle a_\mu\rangle=0$, the holons, $\phi$, will see the same band structure as the electrons in the physical system of interest. 

Considering the example of $t$MoTe$_2$, at sufficient twist angle time-reversal symmetry is broken due to spontaneous Ising ferromagnetism, and a narrow $C=-1$ flat band occurs. If the physical electrons fill half fill this band, then so too do the holons (at mean field level), since they see the same superlattice potential. Because the holons are bosons, they then form an incompressible, \emph{bosonic FQAH state} with the same topological order as the $\nu=-1/2$ bosonic Laughlin state and Hall conductivity, $\sigma^\phi_{xy}=-e^2/2h$. They may be integrated out to yield the FQAH topological quantum field theory (TQFT)~\cite{MacDonald1983,Lu2020,Manjunath2021},
\begin{align}
\mathcal{L}_{\mathrm{\phi}}&=\frac{2}{4\pi}bdb+\frac{1}{2\pi}bd(A-a)+(A_t-a_t)\,\overline\rho\,,
\end{align}
where we have introduced an auxiliary gauge field, $b_\mu$. We define $\overline\rho$ to be the value of the physical charge density, such that the filling is $\overline\rho\times(\textrm{moir\'{e} unit cell area})\equiv\mathfrak{s}=+1/2$. Indeed, the main difference with the usual continuum bosonic Laughlin TQFT is the presence of the final term, which gives the charge ``glued'' to the superlattice and guarantees that the theory makes sense in the absence of an external magnetic field. Notably, in a FQAH or fractional Chern insulator (FCI) state at finite field, $\mathfrak{s}n\in \mathbb{Z}$ should be quantized in units of the topological degeneracy on the torus, $n$~\cite{Lu2020}. 

Assuming that at long wavelengths the spinons can be described with a quadratic dispersion and effective mass, $m_*$, one then obtains a final effective theory for the Chern band system at half-filling,
\begin{align}
\mathcal{L}_{\mathrm{ACFL}}&=\psi^\dagger(i\partial_t+a_t)\,\psi-\frac{1}{2m_*}\left| (i\partial_i+a_i)\psi\right|^2\nonumber\\
&\qquad+\frac{2}{4\pi}bdb+\frac{1}{2\pi}bd(A-a)+(A_t-a_t)\,\overline\rho\,.
\end{align}
Integrating out $b_\mu$ and shifting $a\rightarrow a+A$ gives the effective Lagrangian in Eq.~(3) 
of the main text. However, only the Lagrangian with $b_\mu$ included is gauge invariant on arbitrary manifolds.

We remark that the procedure outlined here holds whether or not the holon FQAH state is chosen to have nontrivial spatial symmetry fractionalization. Indeed, we conjecture that an advantage of the composite fermion framework is that the choice of symmetry fractionalization pattern for the holon FQAH state \emph{determines} the symmetry fractionalization of the Jain sequence states built up by filling Landau levels of composite fermions. We plan to present a more detailed discussion of the composite fermion point of view for spatial symmetry enrichment in future work.


\section{Commensurability oscillations}

A distinguishing feature of the anomalous composite Fermi liquid (ACFL) is that it will display commensurability oscillations as a function of density alone due to its in-built moir\'e periodic potential, while an ordinary Fermi liquid or finite-field CFL will not. In contrast, GaAs displays commensurability oscillations due to an \textit{externally} applied periodical scalar or vector potential~\cite{Kamburov2014} and at finite field. Upon doping away from half-filling, $\bar \rho = \frac12 A_{\text{u.c.}}^{-1}$, the composite fermions feel an effective magnetic field $b_* = 4\pi(\rho_e- \bar \rho)$. Let us assume that the added density (and thus $b_*$) is spatially uniform. We can treat this system as a parabolic band of fermions with effective mass $m^*$ in a magnetic field $b_*$ and a periodic scalar potential. The potential has triangular lattice symmetry with wavevector $Q$ and strength $V_0$. The composite fermions half-fill the lowest band of the moir\'e Brillouin zone.\par 
In this section we discuss the perturbative and semiclassical approaches to commensurability oscillations. While the former approach is well known, discussions of the latter approach are lacking. The advantage of the latter is that it shows that commensurability oscillations naturally arise even in a nonperturbative regime $V_0 \gg \omega_c$; indeed, the ACFL may be in this regime at small $b_*$. \par 
\textit{Perturbative approach. } As noted in Ref.~\cite{Gerhardts1996}, magnetoresistance minima occur at the electron flat band condition. To first order in perturbation theory, the bandwidth of the $n$'th Landau level in a scalar potential of wavevector $Q$ is controlled by the factor 
\begin{equation}
    e^{-Q^2\ell^2/4}L_n(Q^2\ell^2/2) = \frac{\cos(\sqrt{2n}Q\ell - \frac{\pi}{4})}{\sqrt{\pi Q \ell \sqrt{n/2}}} + O(n^{-\frac{3}{4}})
\end{equation}
so at large $n$, using the fact that $n = \rho_e 2\pi \ell^2$ and $k_F = \sqrt{4\pi \rho_e}$, we have the flat band condition $2k_FQ\ell^2 = 2\pi (j-1/4)$. This implies $2R_c/a = j-1/4$ when $a = 2\pi/Q$, while for the triangular lattice this implies $4R_c/\sqrt{3}a =j-1/4$~\cite{Weiss1989}. For composite fermions, this condition becomes $2\sqrt{4\pi(\bar \rho + \Delta \rho_e)} Q / 4\pi \Delta \rho_e = 2\pi(j-1/4)$, which yields Eq.~(11) 
of the main text at large $j$.\par 

\begin{figure}
\centering
\includegraphics[width=0.8\linewidth]{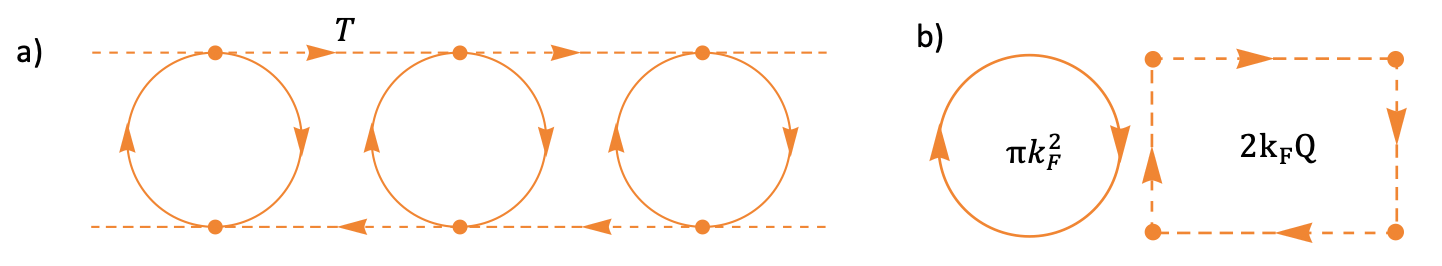}
\caption{\textbf{Network model. } (a) Network model capturing commensurability oscillations due to tunneling $T$ between Fermi surfaces in repeated zone scheme. (b) Relevant $k$-space orbits and areas. }\label{fig:commensurability_network}
\end{figure}

\textit{Semiclassical approach. } 
We follow the semiclassical network model approach to the magnetic density of states as in~\cite{Pippard1962,Pippard1964,Paul2022}. Consider the network model shown in Fig.~\ref{fig:commensurability_network}. For simplicity we are considering a 1D modulation instead of a 2D modulation, since the nature of commensurability oscillations doesn't change qualitatively between the two. We are assuming some phenomenological tunneling $T$ between Fermi surfaces in neighboring zones. The tunneling is best thought of as wavepacket tunneling over a barrier (in $k$-space). The barrier is set by the dispersion, and hence $T$ depends on $V_0$. Although it can occur at other points in the Fermi surface, the tunneling is peaked when the velocity is normal to the barrier; hence as a simplified model we have taken a single hopping between Fermi surface tops and bottoms. 
\par 
To describe scattering at the junctions, we adopt a basis where $(0,1)^T$ and $(1,0)^T$ are the dashed and straight incoming edges, respectively. After a suitable gauge transformation, the scattering unitary at the junction can be written as
\begin{equation}
    U = \begin{pmatrix}
        \sqrt{1-T^2} e^{2\pi i\phi} & -T \\ T & \sqrt{1-T^2}e^{-2\pi i\phi}
    \end{pmatrix}.
\end{equation}
Together with the requirement that electrons pick up phases along links so that the total phase around any closed orbit is equal to the corresponding Aharanov-Bohm flux, this specifies the network model completely. We solve for the spectrum at a given magnetic field $B$ by solving for eigenmodes. Valid eigenmodes must be periodic (up to phase) in the repeated Brillouin zone.\par 

Let $S_0 = \pi k_F^2$ and $S_\square = 2k_F Q$. Note $S_\square$ is not the area of any valid closed orbit, but $S_1 = S_0 + S_\square$ is. Next, note that the network could be considered as 1D array of area-$S_1$ closed orbits, whose mutual overlaps are the area-$S_0$ orbits. These correspond to the ``original" and ``lens" orbits of Ref.~\cite{Paul2022}. It follows from the analysis of Ref.~\cite{Paul2022} that there is a \textit{flat band} whenever $S_0$ and $S_1$ are both quantized, or, equivalently, whenever $S_0$ and $S_\square$ are both quantized. (A flat band in the network model corresponds to an extensive set of localized eigenmodes). In short, flat bands occur when
\begin{align}
    \ell^2 S_0 &=2\pi(n+\gamma)\label{eq:S0}\\
    \ell^2 S_\square &= 2\pi(m-\phi)\label{eq:Ssquare}
\end{align}
where $m,n$ are suitable integers and $\gamma = \frac12$ is a Maslov correction. Using $k_F^2 = 2mE$, Eq. \ref{eq:S0} is $E = \omega_c(n+1/2)$ while Eq.~\ref{eq:Ssquare} is precisely the commensurability condition $2k_F Q \ell^2 = 2\pi(j-\phi)$
for $\phi = 1/4$. The flat band condition coincides with a compressibility maximum and resistance minimum. Note that the all-orders in $V_0/\omega_c$ dependence is encoded in $T$, which does not affect the final result. The agreement of the two approaches supports the robustness of the commensurability oscillations across the relevant range of $b_*$. 

\section{Comment on PH symmetry and the possibility of a Dirac composite fermion}

We wish to remark here on the possibility of particle-hole (PH) symmetry in the ACFL. While PH is present microscopically in the LLL~\cite{Girvin1984} -- leading to proposals of Dirac CFLs~\cite{Son2015,Goldman2018} with an emergent $\pi$ Fermi surface Berry phase -- the Chern band systems we study lack PH symmetry \cite{Reddy2023}, and we anticipate PH asymmetry to occur between states at Chern band fillings $n<1/2$ and $n>1/2$. Nevertheless, it is possible that a PH symmetric response can be an emergent property. In particular, recent work has proposed the emergence of PH and so-called reflection symmetries starting from an HLR description~\cite{Wang2017,Kumar2018,Kumar2019a}, for example on on including the effect of disorder. We leave 
this direction to future work.



%